\documentclass[aps,prb,twocolumn]{revtex4}

\usepackage{graphicx}
\usepackage{amsmath}

\begin{document}

\title{Excitons in solids from time-dependent density-functional theory: \\ Assessing the Tamm-Dancoff approximation}

\author{Young-Moo Byun}
\affiliation{Department of Physics and Astronomy, University of Missouri, Columbia, MO 65211, USA}

\author{Carsten A. Ullrich}
\email[]{ullrichc@missouri.edu}
\affiliation{Department of Physics and Astronomy, University of Missouri, Columbia, MO 65211, USA}

\begin{abstract}
Excitonic effects in solids can be calculated using the Bethe-Salpeter equation (BSE) or the Casida equation of time-dependent density-functional theory (TDDFT). In both methods, the Tamm-Dancoff approximation (TDA), which decouples excitations and de-excitations, is widely used to reduce computational cost. Here, we study the effect of the TDA on exciton binding energies of solids obtained from the Casida equation using long-range corrected (LRC) exchange-correlation kernels. We find that the TDA underestimates TDDFT-LRC exciton binding energies of semiconductors slightly, but those of insulators significantly (i.e., by more than 100\%), and thus it is essential to solve the full Casida equation to describe strongly bound excitons. These findings are relevant in the ongoing search for accurate and efficient TDDFT approaches for excitons.
\end{abstract}


\maketitle

\section{Introduction}

Excitons are bound electron-hole pairs arising in optically excited finite and extended systems. Understanding and predicting excitonic properties is important for the design of novel photovoltaic materials. For example, low exciton binding energies in perovskite solar cells promote the electron-hole separation and thereby enhance power conversion efficiencies.\cite{Miyata2005}

Many-body perturbation theory is a standard method to calculate excitonic properties of solids: one obtains accurate exciton binding energies $E_{\mathrm{b}}$ and optical absorption spectra of semiconductors and insulators by solving the Bethe-Salpeter equation (BSE).\cite{Onida2002} However, the BSE is computationally expensive and cannot be applied to large systems.

Time-dependent density-functional theory (TDDFT) is a computationally cheaper alternative to the BSE,\cite{Ullrich2015}
but its application to the study of excitonic effects in solids is hampered by the
need to approximate the unknown exchange-correlation (xc) kernel $f_{\mathrm{xc}}$. The random phase approximation (RPA) (i.e., $f_{\mathrm{xc}}=0$), the local density approximation (LDA), as well as any standard gradient-corrected semilocal approximation fail to capture excitonic properties of solids due to their inadequate long-range behavior. A very accurate xc kernel can be
derived by reverse-engineering the BSE,\cite{Onida2002,Reining2002}
but it is computationally as expensive. A drastic simplification, known as the long-range corrected (LRC) kernel,
\begin{equation}\label{LRC}
f_{\mathrm{xc}}^{\mathrm{LRC}}=-\frac{\alpha}{\mathbf{q}^2} \:,
\end{equation}
accounts for bound excitons in solids, but it requires a material-dependent parameter $\alpha$, a positive scalar. Inspired by the 
simple form (\ref{LRC}), a whole family of 
LRC-type kernels have been proposed in the literature.\cite{Botti04, Sharma11, Rigamonti15, Berger15, Trevisanutto13}

The performance of LRC-type kernels is typically judged by how well they appear to reproduce experimental optical absorption spectra.
However, a better quantitative measure is the direct calculation of exciton binding energies, which can be achieved by
solving the so-called Casida equation of TDDFT.\cite{Casida1995,Yang2013,Yang2015} This approach is sometimes referred to as
``diagonalizing the excitonic Hamiltonian'', and is formally similar in BSE and TDDFT.
Usually, this is done within the Tamm-Dancoff approximation (TDA), which neglects the coupling between resonant and anti-resonant excitations.
There are some recent studies investigating the performance of the TDA for the BSE;\cite{Sander15,Shao16}
however, the extent to which the TDA affects the solution of the excitonic Casida equation  has not been studied in detail.

In this paper, we assess the  TDA for TDDFT-LRC exciton binding energies of solids. First, we introduce the various LRC-type kernels to be used in this work and examine the effect of the LRC kernel on excitonic properties of solids. Next, we compare LRC exciton binding energies $E_{\mathrm{b}}^{\mathrm{LRC}}$ of solids obtained from the Casida equation within and beyond the TDA. We discover that the TDA has a negligible effect on semiconductors, but a significant effect on insulators. We discuss the origins, practical implications, and limitations of our findings.

\section{Theoretical Background}

\subsection{Dyson equation}

In linear-response TDDFT, there are two ways of calculating optical absorption spectra of periodic systems.\cite{Ullrich2015}
One way is to use the interacting response function $\chi(\mathbf{q}, \omega)$, which is obtained from the Dyson equation
(all quantities are matrices depending on reciprocal lattice vectors $\mathbf{G}$, $\mathbf{G'}$):
\begin{align}
\chi(\mathbf{q}, \omega)&=\chi_{0}(\mathbf{q}, \omega)+\chi_{0}(\mathbf{q}, \omega)\{v(\mathbf{q})+f_{\mathrm{xc}}(\mathbf{q}, \omega)\}\chi(\mathbf{q}, \omega), \nonumber
\end{align}
where $v=v_{0}+\bar{v}=4\pi\delta_{\mathbf{GG'}}/|\mathbf{q}+\mathbf{G}|^{2}$ is the Coulomb interaction, $\chi_{0}$ is the noninteracting response function, and $\mathbf{q}$ is a momentum transfer in the first Brillouin zone (BZ).
$v_{0}$ is the long-range ($\mathbf{G}=0$) part of the Coulomb interaction, and $\bar{v}$ is the Coulomb interaction without the long-range part. In the optical limit ($\mathbf{q} \to 0$), the head ($\mathbf{G}=\mathbf{G'}=0$) of $\chi_{0}$ is given by
\begin{align}
\chi_{0}(\mathbf{q})&=-\frac{4\mathbf{q}^{2}}{(2\pi)^{3}}\sum_{vc}\int_{\mathrm{BZ}}d\mathbf{k}\frac{|\langle c\mathbf{k}|\hat{p}+i[V_{\mathrm{NL}},\hat{r}]|v\mathbf{k}\rangle|^{2}}{(E_{c\mathbf{k}}-E_{v\mathbf{k}})^{3}}, \label{eq:chi0}
\end{align}
where $v$ and $c$ are valence and conduction band indices, respectively, $E_{c,v {\bf k}}$ denotes Kohn-Sham single-particle energies, $\hat{p}$ is the momentum operator, $\hat{r}$ is the position operator, and $V_{\mathrm{NL}}$ is the non-local part of the pseudopotential. The optical spectrum is obtained from the macroscopic dielectric function $\epsilon_{\mathrm{M}}$:
\begin{align}
\epsilon_{\mathrm{M}}(\omega)&=\lim_{\mathbf{q} \to 0} \frac{1}{\epsilon_{\mathbf{G}=\mathbf{G'}=0}^{-1}(\mathbf{q},\omega)} \nonumber \\
&=\lim_{\mathbf{q} \to 0} \frac{1}{1+v_{\mathbf{G}=0}(\mathbf{q})\chi_{\mathbf{G}=\mathbf{G'}=0}(\mathbf{q},\omega)}, \nonumber
\end{align}
where $\epsilon^{-1}$ is the inverse dielectric function. The Dyson-equation approach is computationally relatively cheap, and thus it is the method of choice of most excitonic calculations. However, the method does not allow the precise determination of exciton binding energies,
especially if the excitons are weakly bound.

\subsection{Casida equation}

Alternatively, both optical spectra and exciton binding energies can be obtained from the Casida equation:\cite{Casida1995}
\begin{equation}
 \begin{pmatrix}
  \mathbf{A} & \mathbf{B} \\
  \mathbf{B}^{*} & \mathbf{A}^{*}
 \end{pmatrix}
 \begin{pmatrix}
  X_{n} \\
  Y_{n}
 \end{pmatrix}
 =
 \omega_{n}
 \begin{pmatrix}
  \mathbf{-1} & \mathbf{0} \\
   \mathbf{0} & \mathbf{1}
 \end{pmatrix}
 \begin{pmatrix}
  X_{n} \\
  Y_{n}
 \end{pmatrix}, \label{eq:Casida}
\end{equation}
where $\mathbf{A}$ and $\mathbf{B}$ are excitation and de-excitation matrices, respectively, $X_{n}$ and $Y_{n}$ are $n$th eigenvectors, and $\omega_{n}$ is the $n$th eigenvalue. The matrix elements of $\mathbf{A}$ and $\mathbf{B}$ are
\begin{align}
A^{(v'c'\mathbf{k'})}_{(vc\mathbf{k})}&=(E_{c\mathbf{k}}-E_{v\mathbf{k}})\delta_{vv'}\delta_{cc'}\delta_{\mathbf{k}\mathbf{k'}}+F^{(v'c'\mathbf{k'})}_{\mathrm{Hxc},(vc\mathbf{k})}, \nonumber \\
B^{(v'c'\mathbf{k'})}_{(vc\mathbf{k})}&=F^{(v'c'\mathbf{k'})}_{\mathrm{Hxc},(vc\mathbf{k})}, \nonumber
\end{align}
where $F_{\mathrm{Hxc}}=F_{\mathrm{H}}+F_{\mathrm{xc}}$ is the Hartree-exchange-correlation (Hxc) matrix. In the optical limit, the matrix elements of $F_{\mathrm{Hxc}}$  using the LRC kernel are given by
\begin{align}
F^{(v'c'\mathbf{k'})}_{\mathrm{Hxc},(vc\mathbf{k})}=\frac{2}{V} \Big( \sum_{\mathbf{G}\neq0} \frac{4\pi - \bar{\alpha}}{|\mathbf{G}|^{2}} \langle c\mathbf{k}|e^{i\mathbf{G}\cdot\mathbf{r}}|v\mathbf{k}\rangle \langle v'\mathbf{k'}|e^{-i\mathbf{G}\cdot\mathbf{r}}|c'\mathbf{k'}\rangle \nonumber \\
-\alpha_{0} \frac{\langle c\mathbf{k}|\hat{p}+i[V_{\mathrm{NL}},\hat{r}]|v\mathbf{k}\rangle}{E_{c\mathbf{k}}-E_{v\mathbf{k}}} \frac{\langle v'\mathbf{k'}|\hat{p}-i[V_{\mathrm{NL}},\hat{r}]|c'\mathbf{k'}\rangle}{E_{v'\mathbf{k'}}-E_{c'\mathbf{k'}}} \Big). \label{eq:FHxc}
\end{align}
Here, $V$ is the crystal volume, $\alpha=\alpha_{0} \neq \bar{\alpha}=0$ for $f_{\mathrm{xc}}^{\mathrm{LRC}}=-(\alpha / 4\pi) v_{0}$ (head-only), and $\alpha=\alpha_{0}=\bar{\alpha} \neq 0$ for $f_{\mathrm{xc}}^{\mathrm{LRC}}=-(\alpha / 4\pi) v$ (diagonal). The Casida-equation approach yields very precise exciton binding energies, but is computationally expensive because it requires building and diagonalizing a large matrix.

\subsection{LRC kernel: head-only vs diagonal}

Head-only and diagonal LRC kernels, with $f_{\mathrm{xc}}^{\mathrm{LRC}}=-(\alpha / 4\pi) v_{0}$ and $f_{\mathrm{xc}}^{\mathrm{LRC}}=-(\alpha / 4\pi) v$, respectively, have been used interchangeably because (i) the form of the LRC kernel is not dictated by a rigorous formal derivation, so the two LRC kernels are largely a matter of choice; (ii) the two LRC kernels cause negligible differences in optical spectra of semiconductors such as Si [because $\bar{\alpha} \approx 0.2 \ll 4\pi$ in Eq.~(\ref{eq:FHxc})].\cite{Botti04} However, as we will report elsewhere, we found that the two kernels yield very different results for exciton binding energies of insulators, so it is important to state clearly which version is used. We used the head-only LRC kernel in this work; however, our findings concerning the performance of the TDA hold for both types of LRC kernels.

\subsection{Tamm-Dancoff approximation}

The TDA decouples excitations and de-excitations by setting $\mathbf{B}$ to zero in Eq.~(\ref{eq:Casida}). The TDA is widely used in the BSE and the Casida equation because it cuts the computational cost significantly by reducing the size of the exciton Hamiltonian matrix by a factor of two and changing a non-Hermitian eigenvalue problem to a Hermitian one. However, it turns out that the full Casida equation can be solved at the same computational cost as the TDA,~\cite{Sander15} using a transformation that is well known
from computational chemistry.\cite{Casida1995}  Making use of time-reversal symmetry, Eq.~(\ref{eq:Casida})  can be transformed to a Hermitian eigenvalue equation:
\begin{align}
\mathbf{C}Z_{n}&=\omega_{n}^{2}Z_{n}, \nonumber
\end{align}
where
\begin{align}
\mathbf{C}&=(\mathbf{A}-\mathbf{B})^{1/2}(\mathbf{A}+\mathbf{B})(\mathbf{A}-\mathbf{B})^{1/2}, \nonumber \\
Z_{n}&=(\mathbf{A}-\mathbf{B})^{1/2}(X_{n}-Y_{n}). \nonumber
\end{align}

\subsection{Band-gap corrections: LDA vs scissors shift}

A standard method of producing band structures with the correct band gap is to use so-called scissors operators.
There are many ways of applying the scissors shift to Dyson and Casida equations in Eqs.~(\ref{eq:chi0}) and (\ref{eq:FHxc}) and LRC-type kernels. The scissors shift can be applied to only conduction bands (i.e. replacing $E_{c\mathbf{k}}$ by $E_{c\mathbf{k}}+\Delta$) or to the momentum operator (i.e. replacing $\hat{p}$ by $\{(E_{c\mathbf{k}}+\Delta-E_{v\mathbf{k}})/(E_{c\mathbf{k}}-E_{v\mathbf{k}})\}\hat{p}$) as well, where $\Delta$ is the difference between experimental (or $GW$) and DFT bandgaps.

Due to the many choices involved and the high sensitivity of the LRC kernel, the scissors shift can cause some ambiguities
(we will address these issues elsewhere in more detail). In this paper our focus is on the performance of the TDA; we wish to avoid
any unnecessary distractions and therefore simply work with uncorrected LDA band structures in both Dyson and Casida equations and in the construction of all xc kernels. This impacts the exciton binding energies calculated with and without TDA  in the same way (both are calculated relative to the LDA gap), so a meaningful assessment of the TDA
is possible. On the other hand, to compare optical spectra with experiment, we simply shift them rigidly by  
the difference between the LDA gap and the experimental band gap.

\subsection{Local-field effect}

The local-field effect (LFE) has different meanings in Dyson and Casida equations. In the Dyson equation, the LFE means that $\epsilon^{-1} \neq 1/\epsilon$. The Dyson equation is used to calculate optical spectra and Bootstrap-type kernels, which will be explained in Section IV.A. In the Dyson equation for optical spectra, the LFE is not a matter of choice and should be included. However, in the definition of Bootstrap-type kernels, we 
have the freedom of whether or not to include the LFE, because Bootstrap-type kernels are not constrained by formal derivations.
In the following, we chose to include the LFE when calculating Bootstrap-type kernels to be consistent and focus on the TDA.

In the Casida equation, LFE means that not only the head (i.e. $\mathbf{G}=\mathbf{G'}=0$) term, but also other terms are included in the summation of $F_{\mathrm{Hxc}}$ matrix elements in Eq.~(\ref{eq:FHxc}). In the Casida equation, the LFE is not a matter of choice and should be included.

\section{Computational details}

We used the Abinit code for norm-conserving pseudopotentials, Kohn-Sham eigenvectors and eigenvalues, and $GW$ bandgaps within the LDA.\cite{Gonze09} We wrote our own TDDFT code for calculating exciton binding energies, and used the dp code for optical spectra.\cite{Olevano97} We used experimental lattice parameters and align the optical spectra of GaAs and solid Ne with the experimental band gaps.

In the Dyson equation for optical spectra, we used a 16$\times$16$\times$16 Monkhorst-Pack $k$-point mesh, 4 valence bands, and 20 conduction bands for GaAs and solid Ne. In the Dyson equation for Bootstrap-type kernels, we used a 20$\times$20$\times$20 (20$\times$20$\times$10) $\Gamma$-centered $k$-point mesh, 4 (8) valence bands, 20 (20) conduction bands, and 59 (73) $\mathbf{G}$ vectors for GaAs, $\beta$-GaN, MgO, LiF, solid Ar, and solid Ne ($\alpha$-GaN and AlN). In the Casida equation, we used a 28$\times$28$\times$28 (16$\times$16$\times$16) \{16$\times$16$\times$8\} [8$\times$8$\times$8] $\Gamma$-centered $k$-point mesh, 3 (3) \{6\} [3] valence bands, 2 (6) \{9\} [24] conduction bands, and 59 (59) \{73\} [59] $\mathbf{G}$ vectors for GaAs ($\beta$-GaN and MgO) \{$\alpha$-GaN and AlN\} [LiF, solid Ar, and solid Ne]. Convergence was carefully tested
throughout.

\begin{figure}[b!]
\begin{tabular}{c}
{\includegraphics[trim=0mm 0mm 0mm 0mm, clip, width=0.48\textwidth]{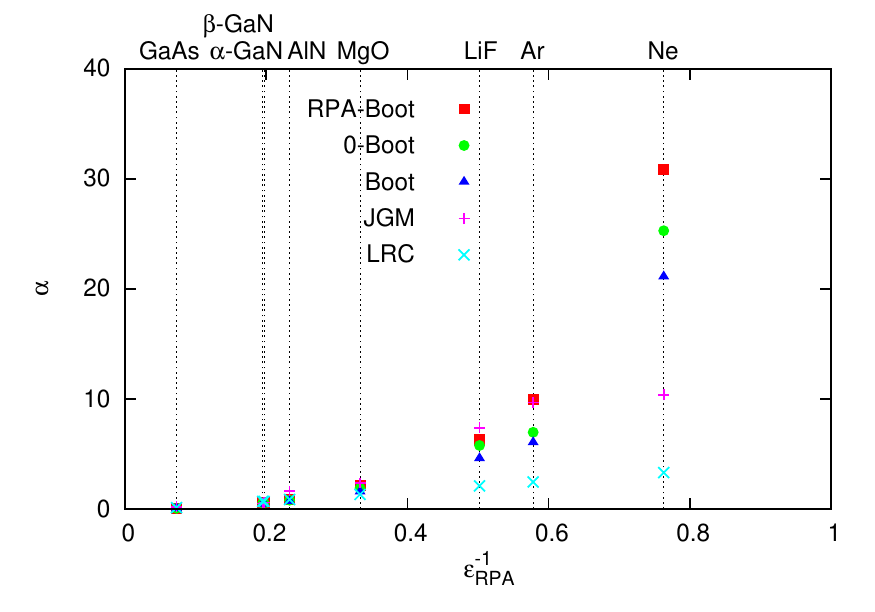}}
\end{tabular}
\caption{(Color online) LRC kernel strengths $\alpha$ [see Eq. (\ref{LRC})] of LRC-type kernels for various materials.}
\label{fig:eps.inv.alpha}
\end{figure}

\section{Results and Discussion}

\subsection{Overview of LRC-type kernels}

We begin by listing five static LRC-type kernels (empirical LRC, Bootstrap, 0-Bootstrap, RPA-Bootstrap, and JGM kernels) which were used in this work.

The empirical LRC kernel ($\alpha = 4.615 \epsilon_{\infty}^{-1} - 0.213$, where $\epsilon_{\infty}$ is the high-frequency dielectric constant) is the first LRC-type kernel for optical spectra of semiconductors.\cite{Botti04} Note that we used the calculated $\epsilon_{\mathrm{RPA}}^{-1}$ instead of experimental $\epsilon_{\infty}^{-1}$; further, $\epsilon_{\mathrm{RPA}}^{-1}$ is greater than $\epsilon_{\infty}^{-1}$ by $\sim$10\%.

The Bootstrap kernel $f_{\mathrm{xc}}^{\mathrm{Boot}}=\epsilon^{-1} / \chi_{0}$, where $\epsilon^{-1}$ is the self-consistent (``bootstrapped'') inverse dielectric function, is a parameter-free kernel for optical spectra of semiconductors and insulators.\cite{Sharma11}

The 0-Bootstrap kernel ($f_{\mathrm{xc}}^{\mathrm{0-Boot}}=\epsilon_{\mathrm{RPA}}^{-1} / \chi_{0}$) is the Bootstrap kernel without
bootstrapping (i.e., only the first cycle of the self-consistent iteration is carried out). Note that $\alpha_{\mathrm{0-Boot}}$ is greater than $\alpha_{\mathrm{Boot}}$ by $\sim$10\% because $\epsilon_{\mathrm{RPA}}^{-1}$ is greater than $\epsilon^{-1}$ by $\sim$10\%.

The RPA-Bootstrap kernel $f_{\mathrm{xc}}^{\mathrm{RPA-Boot}}=\epsilon_{\mathrm{RPA}}^{-1} / \bar{\chi}_{\mathrm{RPA}}$, where $\bar{\chi}_{\mathrm{RPA}}$ is obtained from $\bar{v}$, is a parameter-free kernel for exciton binding energies of insulators.\cite{Rigamonti15} Note that $\alpha_{\mathrm{RPA-Boot}}$ is greater than $\alpha_{\mathrm{0-Boot}}$ by $\sim$10\% because $|\bar{\chi}_{\mathrm{RPA}}|$ is smaller than $|\chi_{0}|$ by $\sim$10\%.

Lastly, the jellium-with-gap-model (JGM) kernel, $\alpha_{\mathrm{JGM}} \approx E_{\mathrm{g}}^{2}$/$n$, where $E_{\mathrm{g}}$ is the band gap and $n$ is the electron density, is a parameter-free kernel for optical spectra of semiconductors and insulators.\cite{Trevisanutto13} Whereas other LRC-type kernels depend on dielectric constants, the JGM kernel depends on band gaps.

 We point out again that we used LDA band gaps for all kernels instead of experimental (or $GW$) band gaps, which affects exciton binding energies of insulators significantly, because our aim is not to test the accuracy of kernels, but to study the effect of the TDA on LRC exciton binding energies. Figure \ref{fig:eps.inv.alpha} shows  the $\alpha$ values of all kernels for different materials. We see that the strength $\alpha$ varies from $\sim$0.1 ($\alpha_{\mathrm{RPA-Boot}}$ for GaAs) to $\sim$30 ($\alpha_{\mathrm{RPA-Boot}}$ for solid Ne).

\begin{figure}[t!]
\begin{tabular}{c}
{\includegraphics[trim=0mm 0mm 0mm 0mm, clip, width=0.48\textwidth]{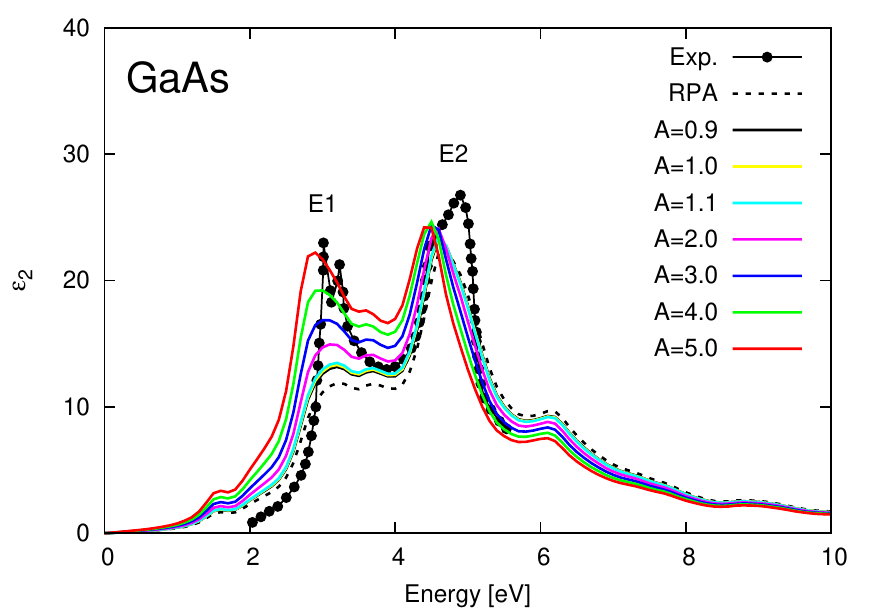}} \\
{\includegraphics[trim=0mm 0mm 0mm 0mm, clip, width=0.48\textwidth]{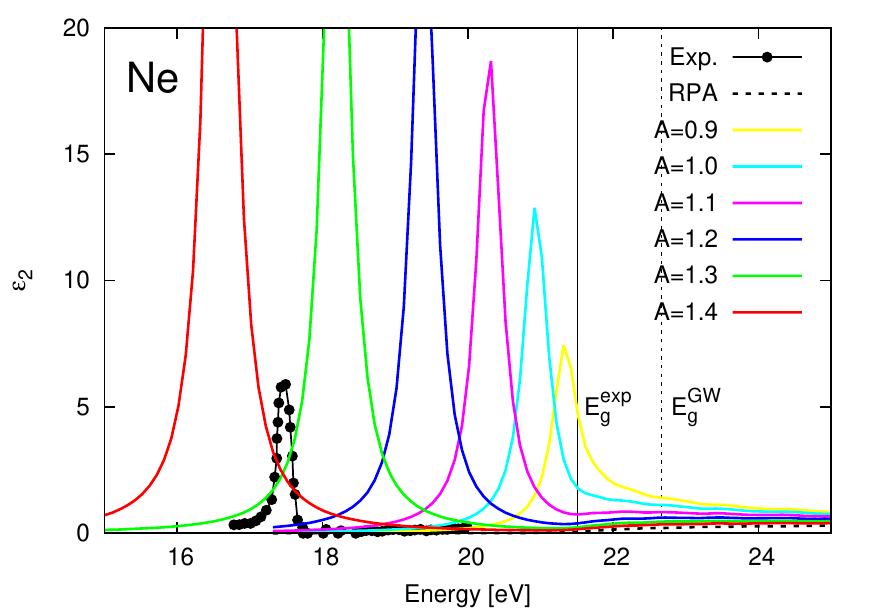}}
\end{tabular}
\caption{(Color online) Experimental and calculated optical absorption spectra of GaAs (top) and solid Ne (bottom). For the LRC kernel, $\alpha=A\alpha_{\mathrm{0-Boot}}$ is used. The spectra are shifted to align the LDA gap with the experimental gap $E_{\mathrm{g}}^{\mathrm{exp}}$; the  $GW$ gap
$E_{\mathrm{g}}^{GW}$ is shown for comparison. A Lorentzian broadening of 0.15~eV (0.2~eV) is used for GaAs (solid Ne).}
\label{fig:optical.spectra.intro}
\end{figure}

\subsection{Effect of the LRC kernel on optical spectra}

Next, we examine the effect of the LRC kernels on optical spectra of solids. Figure \ref{fig:optical.spectra.intro} shows calculated optical spectra of GaAs and solid Ne obtained from the Dyson equation using $f_{\mathrm{xc}}^{\mathrm{LRC}}=-\alpha /\mathbf{q}^2$ ($\alpha=A\alpha_{\mathrm{0-Boot}}$, where $A$ is a scaling factor) and compares them with experimental ones. We chose GaAs and solid Ne because they are extreme examples of semiconductors with weakly bound excitons (Wannier-Mott type) and insulators with strongly bound excitons (Frenkel type), respectively.

There are important differences between semiconductors and insulators. First, exciton binding energies cannot be easily read off the optical spectra of semiconductors since the exciton peaks are too close to the gap, and the binding energies tend to be smaller than the
spectral broadening; by contrast, the binding energies can be quite accurately obtained from the spacings between experimental gaps and excitonic peaks in the optical spectra of insulators.

Second, LRC spectra of semiconductors are insensitive to $\alpha$ (e.g. a 10\% change in $\alpha$ has little effect on the LRC spectrum of GaAs), whereas LRC spectra of insulators are highly sensitive to $\alpha$ (e.g. a 10\% change in $\alpha$ shifts excitonic peaks by $\sim$1 eV in the spectrum of solid Ne). These different effects of the LRC kernel on optical spectra of semiconductors and insulators are important because they are related to different effects of the TDA on LRC exciton binding energies of semiconductors and insulators, which will be shown later. Note that we neglected the effect of the LRC kernel on oscillator strengths or spectral weights (i.e. excitonic peak heights and widths) to focus on excitonic peak positions.

\begin{table*}[t!]
  \caption{Experimental and calculated exciton binding energies (in meV).}
%  \begin{tabular}{ c c c c c c c c c c c c }
  \begin{tabular*}{0.96\textwidth}{ @{\extracolsep{\fill}} r r r r r r r r r r r r }
    \hline \hline
	 & Casida equation  &	GaAs &	$\alpha$-GaN &	$\beta$-GaN &	AlN	&	MgO &	LiF &	Ar &	Ne    \\ \hline
Exp. & 				&	3.27 &			20.4 &			26.0 &	48.0 &	80.0 &	1600 &	1900 &	4080 \\ \hline
%Our Bootstrap &	 No &	5.74 &			23.9 &			23.0 &	153 &	132 &	1470 &	1840 &	3800 \\ \hline
RPA-Bootstrap &	 TDA &	0.334 &			0.927 &			0.875 &	0.00 &	1.72 &	33.3 &	37.7 &	666 \\
0-Bootstrap &	 TDA &	0.285 &			0.811 &			0.720 &	0.00 &	1.43 &	22.4 &	10.8 &	128 \\
Bootstrap &		 TDA &	0.267 &			0.651 &			0.562 &	0.00 &	1.03 &	10.7 &	7.70 &	39.7 \\
JGM &		 	 TDA &	 *  &			0.855 &			0.631 &	0.00 &	2.16 &	93.0 &	33.3 &	6.17 \\
LRC &			 TDA &	0.636 &			1.16 &			1.14 &	0.00 &	0.747 &	1.61 &	1.46 &	1.01 \\ \hline
RPA-Bootstrap &	  Full &	0.344 &			1.06 &			1.01 &	0.00 &	2.12 &	94.7 &	96.0 &	2400 \\
0-Bootstrap &	  Full &	0.293 &			0.919 &			0.829 &	0.00 &	1.72 &	43.2 &	13.7 &	612 \\
Bootstrap &		  Full &	0.278 &			0.735 &			0.649 &	0.00 &	1.20 &	14.8 &	9.14 &	101 \\
JGM &			  Full &	* &			0.971 &			0.727 &	0.00 &	2.77 &	417 &	75.7 &	7.05 \\
LRC &			  Full &	0.670 &			1.33 &			1.32 &	0.00 &	0.855 &	1.89 &	1.54 &	1.06 \\
    \hline \hline
  \end{tabular*}
%  \end{tabular}
  \label{tab:Eb}
\end{table*}

\subsection{TDA and exciton binding energies}

Next, we explore the effect of the TDA on exciton binding energies. Table~\ref{tab:Eb} shows exciton binding energies of different materials obtained from the full and TDA Casida equation using LRC-type kernels. We find that the TDA always underestimates the exciton binding energies. This is consistent with the known fact that  the TDA always overestimates BSE eigenvalues.\cite{Shao16} Secondly, the magnitude of the $E_{\mathrm{b}}$ underestimation by the TDA is small for semiconductors, but large for insulators. For instance, full and TDA $E_{\mathrm{b}}^{\mathrm{LRC}}$ for GaAS differ by 0.034~meV (a 5\% decrease), whereas full and TDA $E_{\mathrm{b}}^{\mathrm{RPA-Boot}}$ of solid Ne differ by 1,734~meV (a 72\% decrease).

There are two possible causes for the large $E_{\mathrm{b}}$ underestimation by the TDA for insulators: (i) large band gaps (e.g. $E_{\mathrm{g}}^{\mathrm{exp}}=1.43$ and 21.5~eV for GaAs and solid Ne, respectively) or (ii) large $\alpha$ values (e.g. $\alpha_{\mathrm{RPA-Boot}}=0.12$ and 31 for GaAs and solid Ne, respectively). The large $E_{\mathrm{b}}$ underestimation by the TDA for insulators vanishes when small $\alpha$ values are used. For example, full and TDA $E_{\mathrm{b}}^{\mathrm{LRC}}$ of solid Ne differ by 0.05~meV (a 5\% decrease) because $\alpha_{\mathrm{LRC}}=3.3$ for solid Ne. This indicates that the large $E_{\mathrm{b}}$ underestimation by the TDA for insulators is solely due to large $\alpha$ values. The large $E_{\mathrm{b}}$ underestimation by the TDA (i.e. a $\sim$50\% decrease) starts to appear when $\alpha \approx 10$.

The general trend is thus that the TDA performs well as long as $E_{\rm b}$ is small compared to the gap (as is the
case for semiconductors), but fails when $E_{\rm b}$ becomes comparable to the gap (as is the case for insulators).
Interestingly, this argument can also be used to rationalize the failure of the TDA to describe plasmons in simple
metals, where the gap is zero.

\subsection{Comparison of Dyson and full Casida equations}

\begin{figure}[b!]
\begin{tabular}{c}
{\includegraphics[trim=0mm 0mm 0mm 0mm, clip, width=0.48\textwidth]{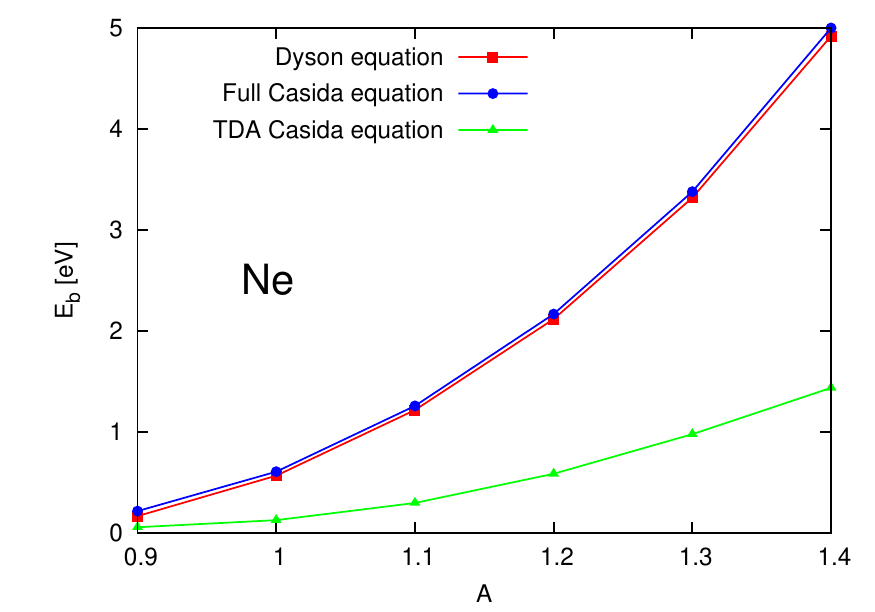}}
\end{tabular}
\caption{(Color online) Calculated exciton binding energies $E_{\mathrm{b}}$ of solid Ne as a function of scaling factor $A$. For the LRC kernel, $\alpha=A\alpha_{\mathrm{0-Boot}}$ is used.}
\label{fig:A.Eb.Ne}
\end{figure}

Next, we verify our finding above from the Casida equation using the Dyson equation. In principle, Dyson and Casida equations are equivalent, so they should result in the same excition binding energy when they use the same kernel. Fig.~\ref{fig:A.Eb.Ne} shows exciton binding energies of solid Ne from the Dyson equation (i.e. from Fig.~\ref{fig:optical.spectra.intro}) and the full and TDA Casida equation as a function of scaling factor $A$.

We find at all $A$ values considered that the Dyson and full Casida equations indeed produce almost identical exciton binding energies, and that the TDA underestimates $E_{\mathrm{b}}^{\mathrm{LRC}}$ of solid Ne by a factor of $\sim$3. This indicates that it is essential to solve the full Casida equation instead of the TDA one when testing whether LRC-type kernels designed for Dyson-equation optical spectra can produce correct and accurate exciton binding energies of insulators.

\subsection{Limitations of our findings}

Finally, we discuss the limitations of our findings. First, our conclusions hold only for LRC-type kernels designed for solids. We did not check the effect of the TDA on other types of kernels that account for bound excitons in solids or are designed for atoms and molecules (some discussion of the TDA in the latter case can be found in Ref. \cite{Ullrich2012}). The large $E_{\mathrm{b}}$ underestimation by the TDA for insulators is partly due to the high sensitivity of the LRC kernel, which is a unique property of the LRC kernel. Hence, it may not occur in non-LRC-type kernels. 

Secondly, we studied only eigenvalues (i.e. exciton energies), not eigenvectors (i.e. exciton states). The impact of the TDA on oscillator strengths in optical spectra and exciton wavefunctions in real space, which are obtained from eigenvectors, remains to be investigated. Third, we studied only the optical limit ($\mathbf{q} \to 0$); the effect of the TDA on finite $\mathbf{q}$ values remains to be tested.\cite{Sander15} Lastly, we studied the  TDA only for excitons in the optical spectra of bulk materials;  in nanoscale systems, additional complications for the TDA
can arise.\cite{Gruning09}

\section{Conclusions}

In summary, we investigated the effect of the TDA on TDDFT-LRC exciton binding energies of solids. We found that the TDA always overestimates LRC eigenvalues and thereby underestimates LRC exciton binding energies. This is consistent with the effect of TDA on $E_{\mathrm{b}}^{\mathrm{BSE}}$. We also found that the magnitude of the $E_{\mathrm{b}}^{\mathrm{LRC}}$ underestimation by the TDA depends on the material: it is negligible for semiconductors with small $\alpha$ values, but significant for insulators with large $\alpha$ values. This behavior of the $E_{\mathrm{b}}^{\mathrm{LRC}}$ underestimation by the TDA is similar to that of the $f_{\mathrm{xc}}^{\mathrm{LRC}}$ sensitivity: LRC excitonic properties of semiconductors are insensitive to $\alpha$, whereas those of insulators are highly sensitive to $\alpha$.

We quantitatively verified that Dyson and full Casida equations produce identical exciton binding energies. This indicates that it is crucial to solve the full Casida equation instead of the TDA one when studying excitonic properties of insulators using LRC-type kernels. 

For now, our conclusions hold only for LRC exciton binding energies of semiconductors and insulators. It will be of interest to study the effect of the TDA for non-LRC-type kernels, and on spectral properties such as oscillator strengths and exciton momentum dispersions.

\begin{acknowledgments}
This work was supported by National Science Foundation Grant No. DMR-1408904.
\end{acknowledgments}

\bibliography{TDA}

\end{document}